\documentclass[aps,pre,showpacs,twocolumn]{revtex4}

\usepackage{graphics}
\usepackage{graphicx}
\usepackage{setspace}
\usepackage{amsmath}
\usepackage{amssymb}

\newcommand{\bvec}[1]{{\mathbf #1}}

\begin{document}

\title{Connecting microscopic simulations with kinetically constrained
models of glasses}

\author{Matthew T. Downton and Malcolm P. Kennett}
\affiliation{ Physics Department, Simon Fraser University, 8888 University Drive, Burnaby, British Columbia, V5A 1S6, Canada}
\date{\today}

\begin{abstract}
Kinetically constrained spin models are known to
exhibit dynamical behavior mimicking that of glass forming systems.  They are often
understood as coarse-grained models of glass formers, in terms of some ``mobility'' field.
The identity of this ``mobility'' field has remained elusive due to the lack
of coarse-graining procedures to obtain these models from a more microscopic point of view.
Here we exhibit a scheme to map the dynamics of a two-dimensional soft disc
glass former onto a kinetically constrained spin model, providing an attempt at 
bridging these two approaches.
\end{abstract}


\maketitle

\section{Introduction}
The origin of the onset of ultra-slow dynamics in glassy systems, and in 
particular, glass-forming liquids, remains a murky subject,
with many competing ideas and tantalizing clues as to underlying 
causes, despite 
years of effort by a large community of 
researchers \cite{Classic}. 
Recently it has become increasingly clear that dynamical 
heterogeneities, 
regions of atypically  fast dynamics that are localized in space and time, 
are intimately connected to the phenomenon of glassiness 
\cite{Sillescu,Ediger,Glotzer,Richert,Castillo}, becoming 
increasingly important at lower temperature scales towards and below the glass 
transition temperature $T_g$  \cite{DHexp,Kob,Donati1,Donati2}.

Early ideas about heterogeneous dynamics focused on the
idea of co-operatively rearranging regions which grow with 
 decreasing temperature \cite{Adam}. Currently, molecular dynamics (MD) simulations
of supercooled liquids allow much greater access to the microscopic 
details of this heterogeneity
\cite{HarrowellDH,Yamamoto,Andersen}.  This has
included the observation of ``caging'' of particles, and string-like excitations that 
allow particles to escape these cages  \cite{Donati1,Perera2,Gebremichael1,Stevenson}, 
which has been confirmed in experiments on colloidal glasses \cite{Kegel-Weeks}.
However, MD simulations have the drawback that it is difficult to reach the
 low temperatures and long times characteristic of the glassy phase.  

An alternative approach to reach low temperatures and long times,
 but without a microscopic foundation, is to study
simple models of glassiness, kinetically constrained models (KCMs) 
\cite{Ritort,FA,Graham,CG,PNAS,LMSBG},
such as the Fredrickson-Andersen (FA) model \cite{FA} or the East model \cite{Evans}, 
or variations such as the North-East model \cite{PNAS},
 which mimic the constrained dynamics of real glassy systems but have trivial thermodynamics.  
These may be viewed as effective models for
glasses, in terms of some coarse-grained degree of freedom often labelled ``spins'', 
also termed a ``mobility field'' by some authors \cite{PNAS}. 
In Fredrickson and Andersen's original work, they posited that the degrees of freedom may be
high and low density regions, related to earlier suggestions
by Angell and Rao \cite{ARao}.  Despite these appealing physical pictures, it has not been 
particularly clear to what physical quantity this ``mobility field'' corresponds. 
 If KCMs are truly
effective models of glassy behavior then it should be possible to make a connection
between some set of degrees of freedom, 
 in a MD simulation, for instance, and a KCM.

In this paper we propose a specific coarse-graining procedure to explore 
whether a link can be made between MD 
simulations and KCMs of glassiness. Previous work in this direction  
found evidence of dynamic facilitation in MD simulations \cite{Donati2,Vogel},
however, there was no attempt to map the dynamic facilitation onto a KCM.
We use an approach directly related to the
idea of a ``mobility'' field, using the local mean-square displacement (MSD) in a suitably
defined box to define a spin variable.  Regions with large average MSD correspond to ``up''
spins and those with low average MSD correspond to ``down'' spins.  We give specific
details of our procedure below. 

We investigate the time and
length-scale dependence of this coarse-graining.  The two characteristic time scales are
the beta relaxation time scale, $t_\beta$ (corresponding 
physically to the time for relaxation within a cage) and the longer alpha relaxation 
time scale, $t_\alpha$, which  corresponds to the time-scale on which structural relaxation
of cages occurs.  We find that evidence of dynamic facilitation becomes much stronger at
longer times of order $t_\alpha$, than at earlier times of order $t_\beta$.
We also study the effect of changing the size of the coarse-graining box, $l$, 
in space and consider values $0.02 \leq l/L \lesssim 0.25$, where $L$ is the system size.
With appropriate choices of time and lengthscales, we find a clear mapping from our MD
simulations onto a KCM similar to the 1-spin facilitated FA model.  
This is not what one might naively expect.  Since we study a fragile glass-former, we
expect to find a KCM which exhibits super-Arrhenius relaxation, whereas the 1-spin 
facilitated FA model has Arrhenius relaxation -- the super-Arrhenius growth of timescales
is absorbed into the coarse graining time, which is most effective at capturing
kinetically constrained behaviour when it is of order $t_\alpha$ contrary to
expectations based on the idea of a mobility field \cite{FA,PNAS}.

The demonstration of a coarse-graining procedure to translate from a microscopic model
to a coarse-grained KCM for glasses can shed light on the following: 
it provides a physical interpretation to the ``mobility field''; 
it can give a stronger theoretical justification for the use of KCMs
to study glassy dynamics; and may open the door to further exploration of the link 
between microscopic models and long-time features of dynamics, i.e. answering the 
question: for a given interparticle interaction potential, how will the dynamics of 
the glass behave?

This paper is structured as follows: in Sec.~\ref{sec:MD} we give details of our molecular
dynamics simulations, in Sec.~\ref{sec:CG} we describe our coarse-graining procedure and
present the results of coarse-graining MD simulations, and then in Sec.~\ref{sec:Disc} we
discuss our results.

\section{Molecular Dynamics Simulations}
\label{sec:MD}
We study the dynamics of a well-characterized glass former: the
binary soft disc model with a potential of the form 
$\epsilon\left(\frac{\sigma_{\alpha\beta}}{r}\right)^{12}$ in two dimensions 
\cite{Perera,Yamamoto,Perera2}.  This mixture of discs with size ratio 1:1.4
inhibits crystallization upon cooling.  We choose a 75:25 ratio of small to
large particles and cool the system with the density fixed at $\rho=0.85 
\sigma_{11}^{-2}$. All
temperatures and lengths are quoted in the standard reduced units of the Lennard-Jones
potential using the small disc diameter $\sigma_{11}$, as a length scale.  This mixture has the same glassy 
characteristics as the model previously documented in Ref.~\cite{Perera}.
Post-equilibration calculations are performed using the recently introduced 
iso-configurational (IC) ensemble \cite{AWC1}.  Our results are qualitatively similar
if we follow a single trajectory with no IC averaging, 
but IC averaging gives smoother trends as a function of temperature.

\begin{figure}[htb]
 \includegraphics[width=0.7\columnwidth]{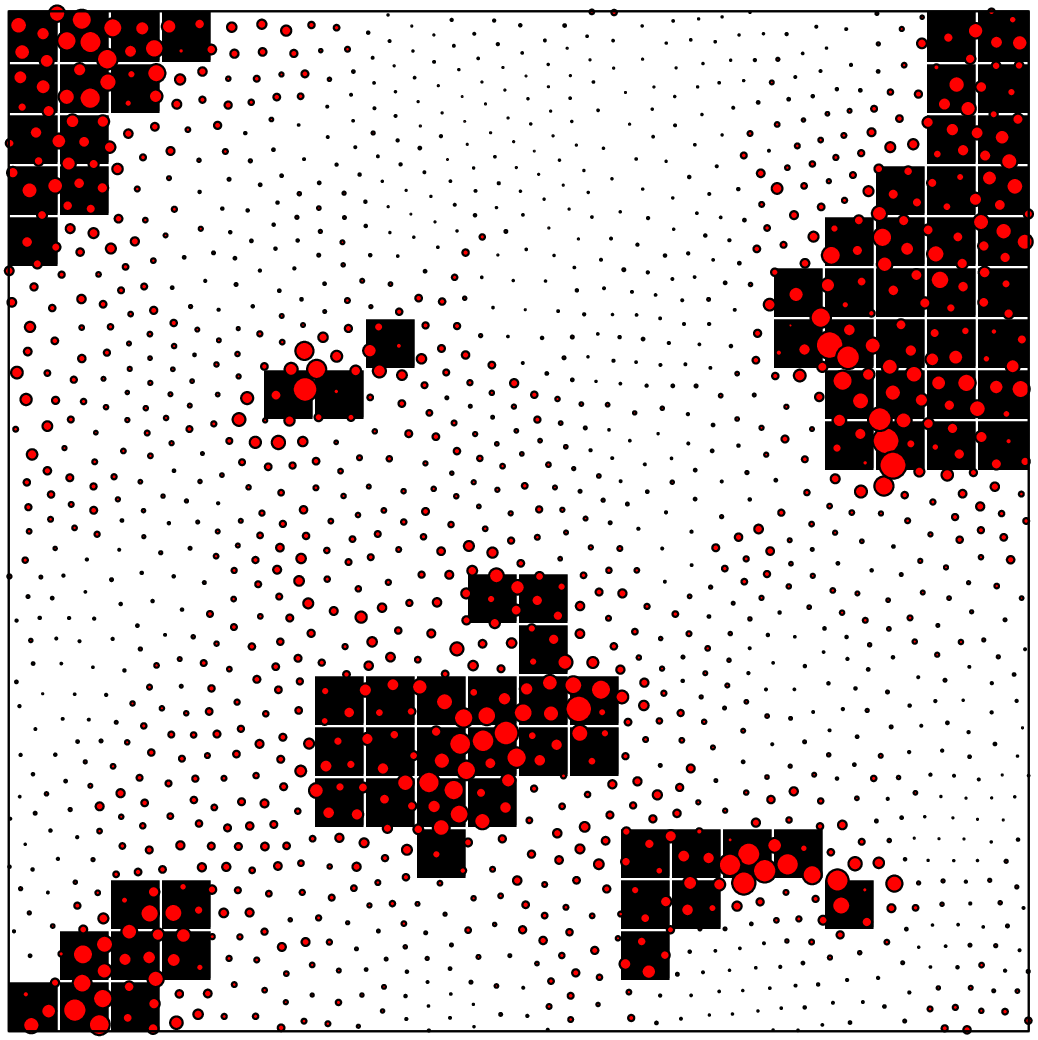}
\caption{Averaged mean square displacement from IC simulations of length
$\tau_e$ at $T=0.360$. On this time scale there is movement throughout the cell.Not all of the the larger groups grow into regions of large displacements at
longer timescales. We show the corresponding spin configuration, with up spins
represented as black squares.}
\label{fig:pnas}
\end{figure}

In Fig.~\ref{fig:pnas} we show the MSD for each particle averaged over 
500 independent trajectories started from the same particle configuration 
in a $N = 1600$ particle system.  Each trajectory is evolved for a time 
$\tau_e$, where $\tau_e$ is the time that it takes for the self-intermediate 
scattering function for the small particles $F_{s}(k,t)=\frac{1}{N}\sum_{i}
\left<\sin\left(k\left|r_{i}(t)-r_{i}(0)\right|\right)/
\left(k\left|r_{i}(t)-r_{i}(0)\right|\right)\right>$ 
(note that the form we use has already been averaged over angle)
to decay by a factor of $1/e$ -- this is roughly $t_\alpha$.  The
$k$ value chosen is that of the first peak of the static structure factor. 
It is clear that there are regions of much  higher MSD
than the average, and that these regions are reasonably widely spaced.
The important question for mapping the dynamics to a KCM is how such
regions influence the behavior of their neighbors.

The MSD in the IC ensemble simulations can be seen as a 
measurement of the propensity for a particle to move based on the initial configuration. 
As noted by Widmer-Cooper and Harrowell, each trajectory within the
ensemble does not reproduce the same dynamics \cite{AWC1}. The final
propensity is therefore the composite of a set of
trajectories that is determined solely by the initial particle positions.  
We can follow the change of the propensity in time by following a
\textit{single} trajectory and performing IC simulations separated by a time $\tau_{s}$
(which we mostly take to be $\sim \tau_e$).  
The KCM that we determine is one that is obtained from an IC average over 500 trajectories.

\section{Coarse graining procedure}
\label{sec:CG}

There are two parts to our coarse-graining procedure.  First, we identify 
the spins that enter in the KCM (using the results of the MD
simulations).  Second we infer the dynamics of these spins.
Specificly, we construct a model which has a Hamiltonian 

\begin{equation}
H = \frac{J}{2}\sum_i s_i,
\label{eq:model}
\end{equation}
where $s_i$ is a ``spin'' variable on a site $i \in [1,N_S]$ (where $N_S$ is the number of
sites in the spin model) for which up ($s_i = 1$) corresponds to an
active region and down ($s_i = -1$) corresponds to an inactive region, and $J$ is some 
(yet to be determined) energy scale. The first part of the coarse-graining procedure
is to find a way to determine the separation of regions into up and down spins.
 In general one might also consider
terms in the Hamiltonian related to spin-spin interactions \cite{LMSBG},
but in their simplest forms, KCMs are usually 
taken to have the form in Eq.~(\ref{eq:model}).
This implies that at high temperatures there are no static correlations, as
one expects in a liquid.
The model Eq.~(\ref{eq:model}) has no interactions
and any glassy phenomenology must come from the dynamical rules that govern 
how spins flip.  These dynamical rules are usually stated in the form that
the probability of a spin flipping is dependent on the state of its
neighbors \cite{FA}. To be more precise, we can note that Glauber rates for flipping 
a spin $i$ are given by \cite{Ritort,LMSBG}:

\begin{eqnarray}
w_i(\bvec{s}) = f_i(\bvec{s}) \left\{ \begin{array}{cc} n_\downarrow , & s_i = 1 \\
n_\uparrow, & s_i = -1 \end{array} \right.
\end{eqnarray}
which respects detailed balance, and the concentration of up spins (with $n_\uparrow + 
n_\downarrow = 1$) is

\begin{equation}
n_\uparrow  = \frac{1}{1 + e^{J/T}},
\label{eq:nup}
\end{equation}
and $\bvec{s} = (s_1, \ldots, s_{N_S})$.
We determine the function $f_i(\bvec{s})$ assuming that it has the form
$f(m)$, where $m$ is the number of up spins on sites neighboring site $i$,
similarly to the formulation of the FA model \cite{FA}.
We analyze the data from the MD simulations to determine $f(m)$.
It is desirable that the results be relatively insensitive to the 
parameters entering the coarse-graining procedure, which is what we find.

Our coarse-graining procedure to determine spins and sites 
is as follows: we perform a set of simulations 
to give $n \simeq 100$ timesteps in the IC ensemble.  The timesteps are
chosen to be $t_\beta$, $0.6\tau_{e}$, and $\tau_e$ to check the coarse-graining in time.
We find that the fitting form $\tau_e = \frac{0.85}{T} e^{\left(\frac{0.5}{T}\right)^{4.5}}$ 
works well over the entire temperature range we consider ($T = 0.36$ to $T = 0.96$), 
although for $T \leq 0.48$ the form $\tau_e = 0.025 e^{\left(\frac{1.1}{T}\right)^2}$ 
works equally well.
We take each of the snapshots of IC averaged particle configurations 
and coarse-grain in space, by dividing the
sample into $(l/L)^2$ boxes lying on a square lattice \cite{footnote}, so that each 
particle is assigned to a box. 
We take $l$ to be small and fixed to $l=2$ most of the time (this appears to be roughly the length-scale
of the cage-breaking process, and also of the order of the dynamic
correlation length, $\xi$ \cite{Perera}) and assume $l$ to be temperature and 
time independent.  
Since we have at most 1600 particles in our
system, when we go to large coarse graining lengths ($l \geq 8$), we 
start to get close to the system size and finite size effects are
important, i.e. $l/L \gtrsim 0.2$.

\begin{figure}[htb]
 \includegraphics[width=0.7\columnwidth]{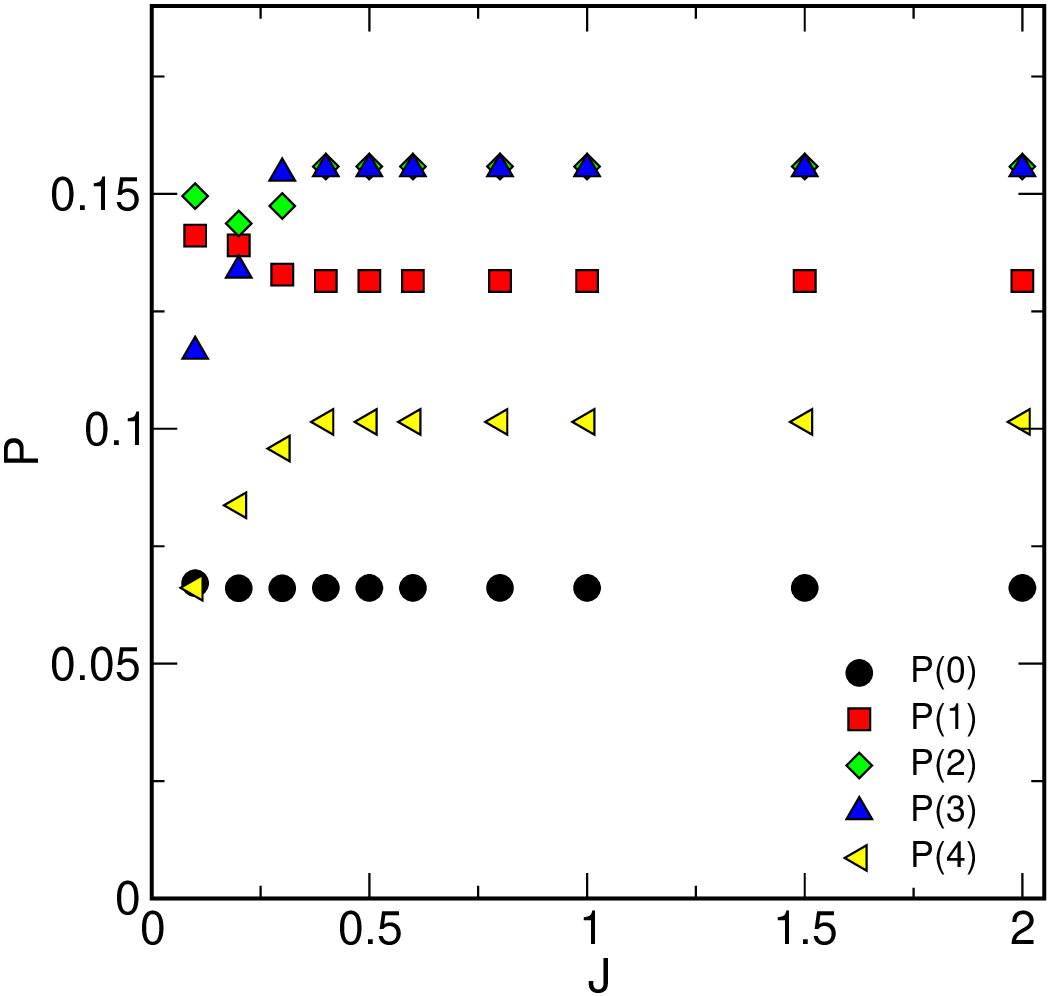}
\caption{$P(m)$ at $m =$ 0, 1, 2, 3, and 4,
for spin flips down to up as $J$ is varied, with a
coarse-graining timescale of $\tau_e$ at $T = 0.36$ with $l=2$.}
\label{fig:checks}
\end{figure}

We associate a spin with each box, either up or down depending on whether the
MSD per particle in the box is larger or smaller than some cutoff.  
We adjust the cutoff so that $n_\uparrow$ takes its equilibrium value, Eq.~(\ref{eq:nup}).
This leaves the freedom to choose the energy scale $J$.
 A seemingly natural energy/temperature
scale associated with glassy dynamics appears to be that where the 
relaxation time for the small particles starts to stretch more quickly and there is
a marked onset of dynamic heterogeneity;
roughly $T \sim 0.5$.  This is also the temperature where
the scaling between diffusion and relaxation times changes \cite{Perera},
motivating us to choose  $J = 0.5$ (in the same units as $T$).
However, we check that our results are robust under varying this choice 
(see Fig.~\ref{fig:checks}). 
In general we find that if $J \gtrsim 0.3$ the
probabilities of spin flips that we identify are identical.

Now, one of the assumptions that underlies writing down Eq.~(\ref{eq:model}) is that 
there are no static correlations.  Given that the spins we consider are themselves
defined through dynamics, albeit within a single coarse-graining time, 
it is difficult to be certain that one can extract truly
static correlations.  Nevertheless, we checked for static correlations in the spin model
defined above and find that at high temperatures there are no static correlations, whereas 
for temperatures below about $T = 0.45$, there are  ``static'' correlations on a lengthscale of 
up to two lattice spacings (at $T = 0.36$), that appears to be growing with decreasing 
temperature.  This is in accord with the expectation that as the spins diffuse, they
lead to local relaxations that appear as a static correlation in a single snapshot, but
can be resolved, for example, by the dynamic four-point susceptibility
that has been discussed extensively in the 
literature \cite{Dasgupta,Donati3,Doliwa,FranzParisi,Glotzer2,BB,Toninelli,Berthier}.
It is also possible that extra terms involving spin-spin interactions 
should perhaps be included in the model as these can also lead to static correlations, 
but the relatively short correlation length, and the existence of dynamic correlations as
discussed above suggests that we can ignore these interactions as a first approximation.  
We shall proceed under this assumption of non-interacting spins and discuss some consequences of 
interactions that appear to give small corrections to our non-interacting results.

\begin{figure}[htb]
  \includegraphics[width=0.7\columnwidth]{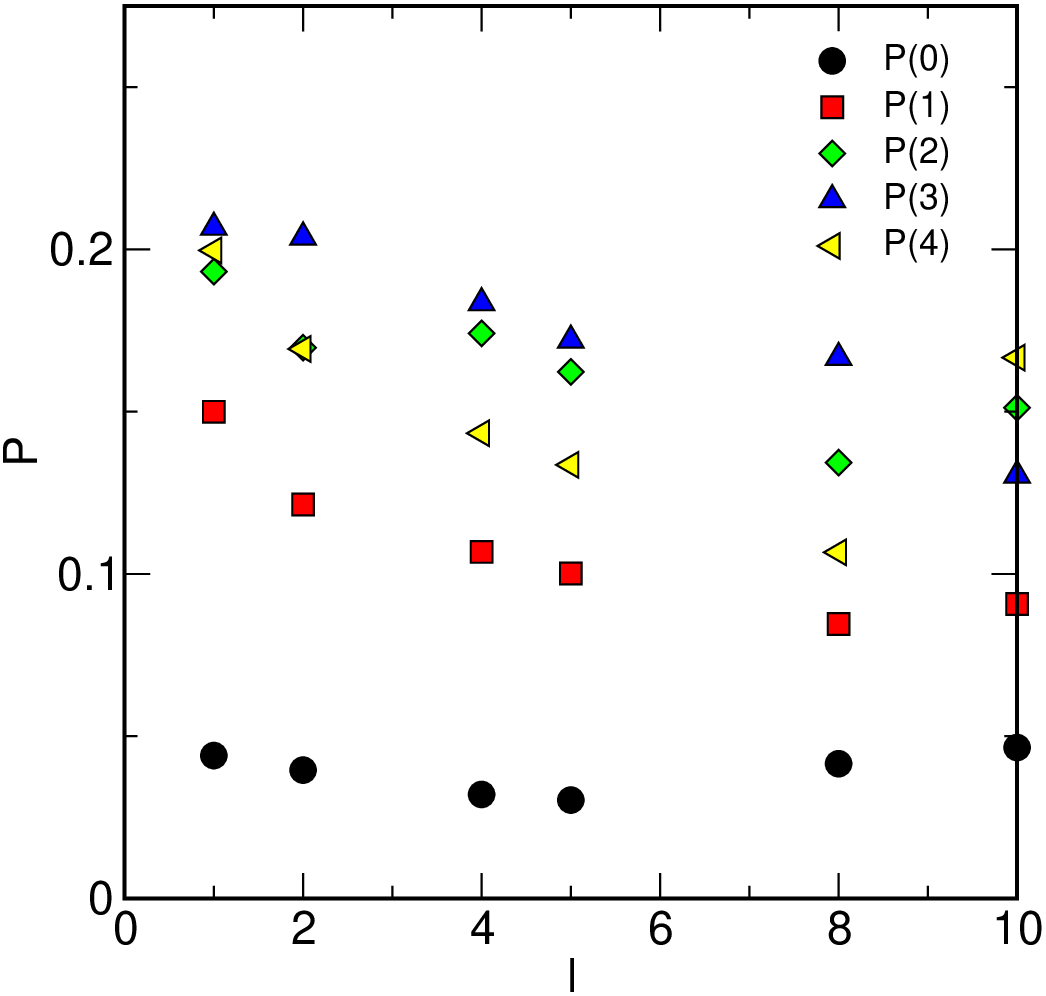}
\caption{$P(m)$ as a function of $m$
for spin flips up  to down as $l$ is varied, with a
coarse-graining timescale of $\tau_e$ at $T = 0.36$ with $J=0.5$.}
\label{fig:checks2}
\end{figure}

 In Figs.~\ref{fig:checks} and \ref{fig:checks2} we show some of our
checks on the coarse-graining procedure.  All of these are at $T = 0.36$.
In particular, in Fig.~\ref{fig:checks} we show how $P(m)$, the probability 
of a spin flip between consecutive time steps changes with variations in $J$ 
at fixed $l = 2$ and $T$ for spin flips from down to up. [At a fixed temperature,
 $P(m) \propto f(m)$.]  In Fig.~\ref{fig:checks2}
we show how $P(m)$ changes with variations in $l$ at fixed $J$ and $T$ for spin
flips with up to down.  Comparable results are found for the spin flips not
shown. Statistical error bars are comparable to the size of the symbols.

\begin{figure}[htb]
 \includegraphics[width=0.7\columnwidth]{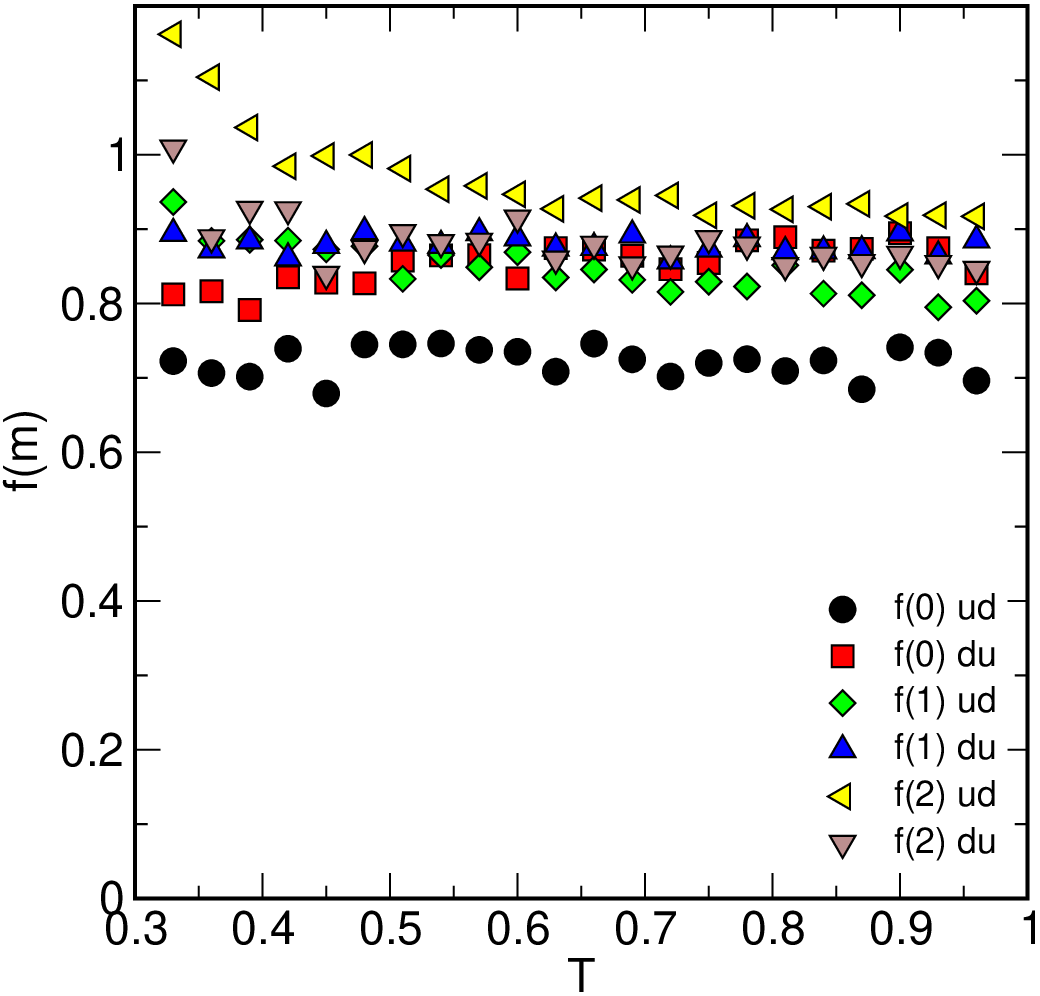}
 \includegraphics[width=0.7\columnwidth]{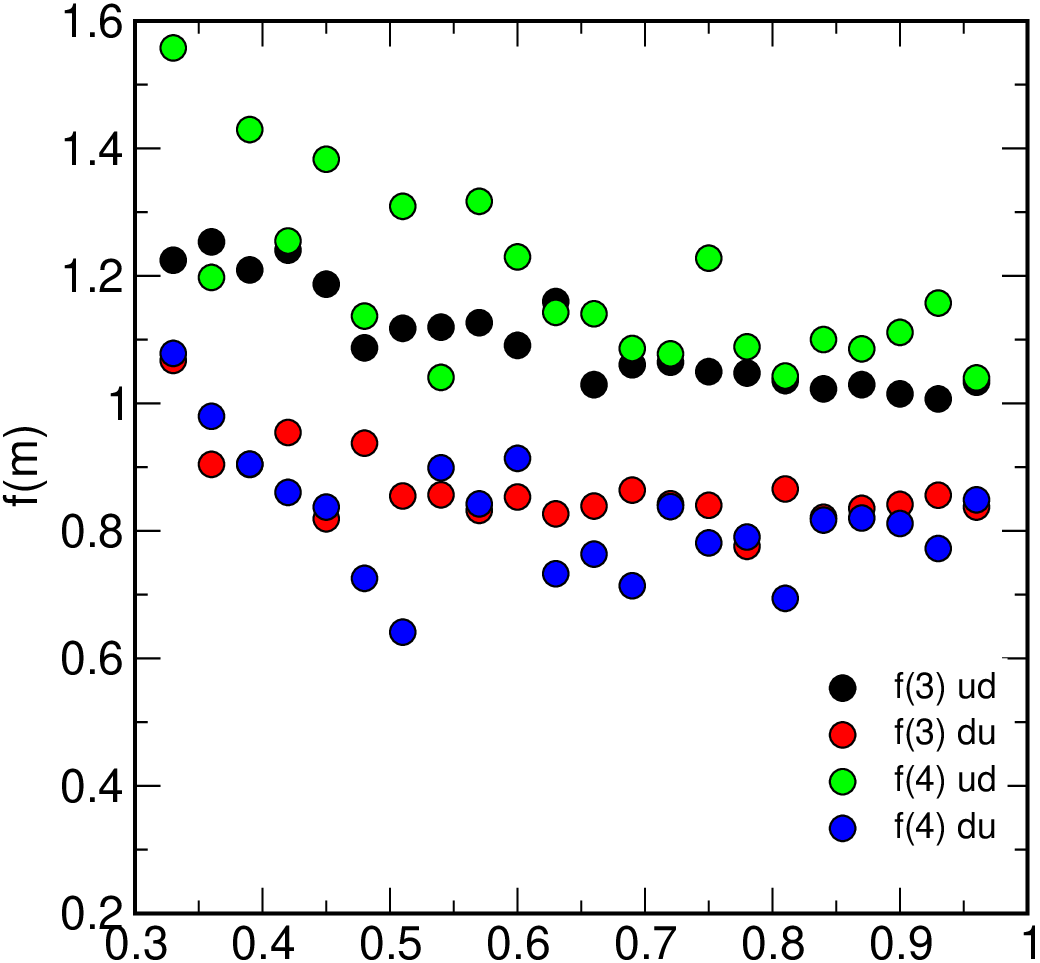}
\caption{$f(m)$ as a function of temperature from both up to down and
down to up spin flips
a) m = 0, 1, 2; b) m = 3, 4 with coarse graining time $t_\beta$}
\label{fig:intercept}
\end{figure}

\begin{figure}[htb]
 \includegraphics[width=0.7\columnwidth]{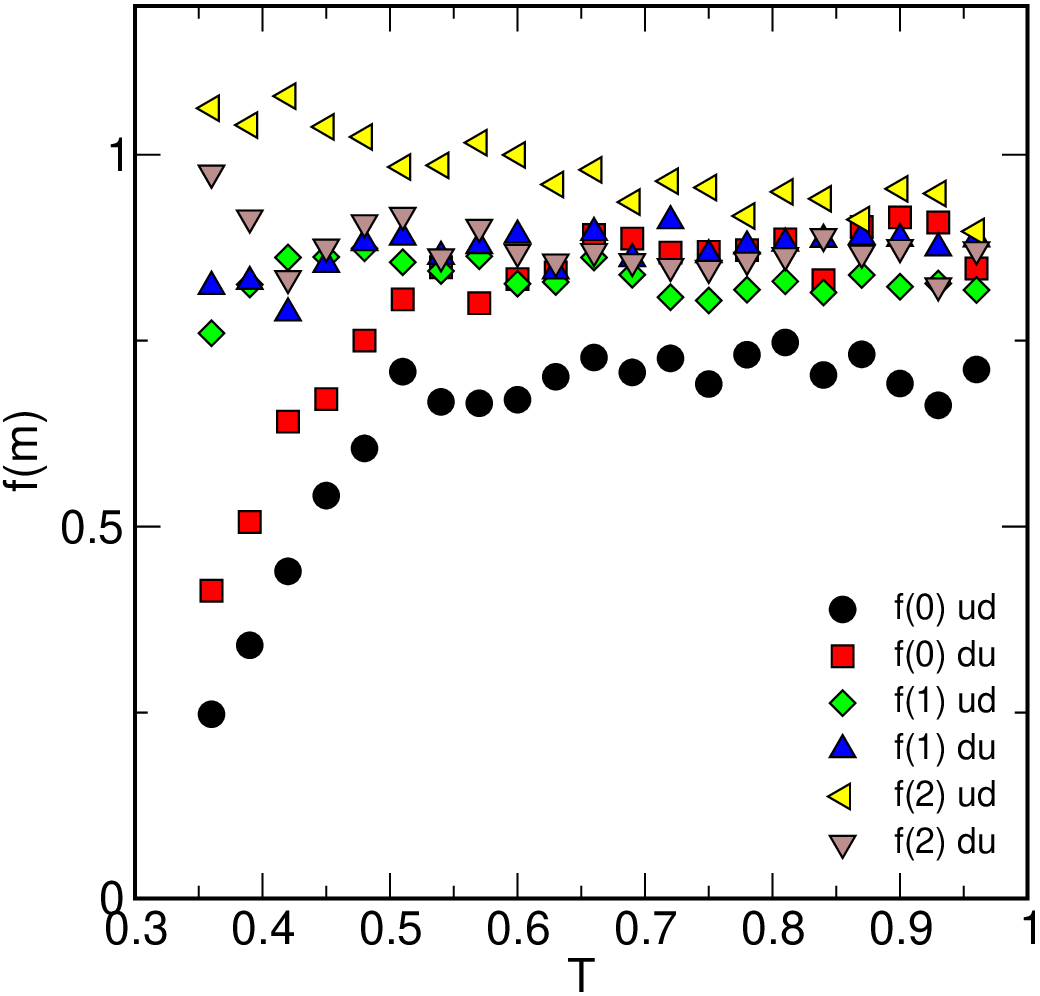}
 \includegraphics[width=0.7\columnwidth]{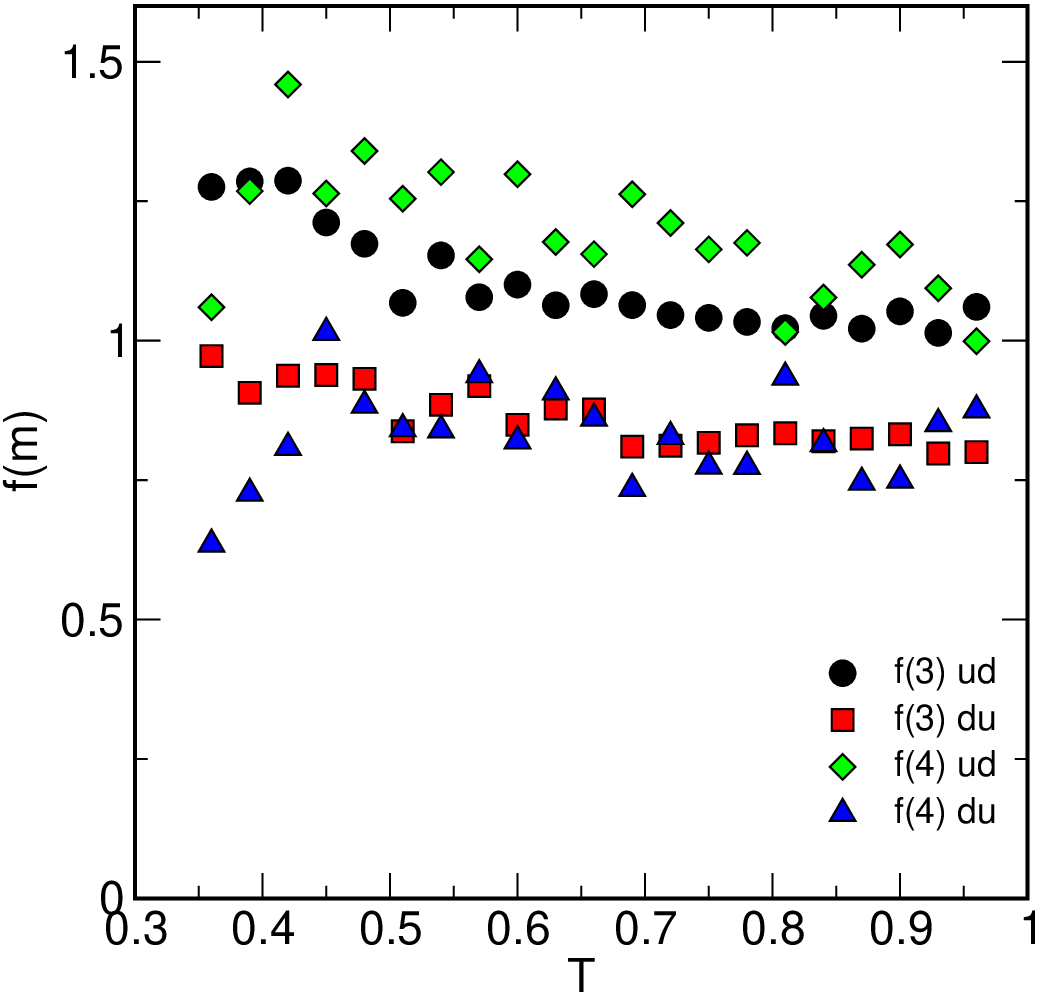}
\caption{$f(m)$ as a function of temperature from both up to down and
down to up spin flips
a) m = 0, 1, 2; b) m = 3, 4 with coarse graining time $\tau_e$}
\label{fig:intercept2}
\end{figure}

We have thus defined our spins.  Now we must understand their dynamics.
To do this, we ask the question, for an up or down spin with $m$ nearest 
neighbors that are up spins, what is the probability that it will flip in a 
given time-step?  This is the way that the classic FA model is posed.
We display the function $f(m)$ as a function of temperature in Figs.~\ref{fig:intercept}
and \ref{fig:intercept2} for coarse graining times of $t_\beta$ and $\tau_e$
respectively.

In Figs.~\ref{fig:intercept}a) and \ref{fig:intercept2}a) we consider $m =$ 0, 1, and 2, 
whilst for clarity, in Figs.~\ref{fig:intercept}b) and \ref{fig:intercept2}b) 
we show $m =3$ and $m=4$.  The behaviour between these two coarse-graining times is 
quite distinct.  For a coarse-graining time of $t_\beta$ there is no strong tendency
towards kinetically constrained dynamics, whereas for a coarse graining time of
$\tau_e$ there are quite strong indications.
For a coarse graining time of $\tau_e$, at low temperatures ($T \lesssim 0.5$), $f(0)$, 
$f(1)$ and $f(2)$ are distinctly different, and there is qualitative agreement between
 $f(m)$ determined from either up to down spin flips or down to up spin flips.
Similar results are seen for $f(3)$ and $f(4)$.

Detailed balance implies that $f(m)$ determined from either type of spin
flip should be the same if the system is described precisely by a model of the type in
Eq.~(\ref{eq:model}).  In Figs.~\ref{fig:intercept} and \ref{fig:intercept2} there are
small but clear differences between $f(m)$ determined from the two types of spin flips.
We believe that there are two sources for this discrepancy.  Probably most important is
the presence of spin-spin interactions which are not accounted for in Eq.~(\ref{eq:model}).
Such interactions mean that $f(m)$ does not depend solely on $m$, although at
low temperatures it is clear that $m$ is the most important variable controlling
the behaviour of $f(m)$.
Secondly, the magnitude of the discrepancy between rates varies from temperature
to temperature point.  This is likely to be from biases that are forced on us by 
computational restrictions.  In principle, one would like to equilibrate a large number of
independent particle configurations, then perform an IC average for each initial condition 
to determine $f(m)$.  In practice, it is computationally
expensive to equilibrate at low temperatures, so we equilibrate one particle 
configuration and then use this to generate subsequent particle configurations
from one member of the IC ensemble.  This introduces a sampling bias
that is likely to contribute to the discrepancy between the rates.  
We note that even for a single trajectory,
there are signatures of kinetically constrained dynamics, but cleaner 
results are obtained from our IC averaging, and we expect averages over more
independent particle configurations to yield more precise results.
Despite these caveats, Fig.~\ref{fig:intercept2} clearly illustrates that the 
coarse-graining procedure we have devised gives strong evidence of kinetically
constrained dynamics, and appears to give a mapping of a microscopic simulation to a KCM.  

The signatures of kinetically constrained behavior only start to show up in the
coarse grained spin model for $T \lesssim 0.5$.
which is around the temperature at which the relaxation time for small particles
increases quickly with decreasing temperature, and is also around where
the Stokes-Einstein relation breaks down. 
An important point to note about KCMs is that they are
expected to give the best account of glassy dynamics when $n_\uparrow \ll 1$. 
In the temperature range we consider, this is only approximately true, for instance, 
at $T = 0.36$, $n_\uparrow \simeq 0.2$ (with $J = 0.5$, although 
for larger $J$, $n_\uparrow$ is considerably smaller, and the $f(m)$ 
we have determined are unchanged).  The requirement of a vanishing number of up spins
suggests that to extract a specific KCM from our data, one should consider the
limit that $T \to 0$.  Figure~\ref{fig:intercept2} certainly suggests that as 
$T \to 0$, $f(0) \to 0$, 
but it is harder to determine the fate of $f(m)$ for $m\neq 0$.  However, our
results suggest a KCM that might apply in the limit $T \to 0$ defined by
(with $\alpha$ constant)
\begin{eqnarray}
f(m) = \left\{ \begin{array}{cc} 0, & m =0 \\ \alpha, & m \geq 1.
\end{array} \right.
\label{eq:finalKCM}
\end{eqnarray}
The most important feature of the model that is realised, whether it is exactly as
in Eq.~(\ref{eq:finalKCM}) or not, is that it is of the one-spin facilitated type
\cite{Ritort}.  It appears unlikely from Fig.~\ref{fig:intercept2} that 
$f(1) \to 0$ as $T \to 0$ as would be required for a two-spin facilitated KCM.
We have verified with Monte Carlo simulations that this model gives timescales 
that diverge in an Arrhenius fashion at low temperatures.  However, the fact that 
the timestep in the KCM is also strongly temperature dependent (through $\tau_e$)
leads to a fragile behaviour of timescales when expressed in terms of the time units
of the MD simulations, i.e. $\tau \sim e^{A/T^2}$ for some constant $A$.

\section{Discussion}
\label{sec:Disc}

There are several goals in trying to map MD simulations of a glass former onto a 
KCM.  The first is making a connection between microscopic particle motions and
some effective theory of glassy dynamics.  A second is to determine the long-time 
dynamics of a given glass former at very
low temperatures where MD simulations are ineffective, say through Monte Carlo simulations of the KCM,
or in some cases, analytic calculations. 

The mapping of MD simulations to a KCM that we have achieved is not what
one might naively expect.  From the discussions in the literature \cite{PNAS,FA},
the expectation would be that for coarse-graining on some small lengthscale
(as we do), and on a timescale much less than $t_\alpha$ or $\tau_e$, one
obtains a KCM which has within it the physics of the alpha relaxation time.
Our attempts along these lines are shown in Fig.~\ref{fig:intercept}, which
clearly indicates that this expected behaviour does not hold.  It is only when
one coarse-grains on timescales of order $t_\alpha$ that we start to see effective
kinetic constraints emerging.
In order to get fragile glass like behaviour, as seen in the MD simulations, from
the constant coarse-graining time scenario, one would require the KCM obtained 
from the coarse-graining procedure to 
be multi-spin facilitated, since single-spin facilitated models of the type found in 
Eq.~(\ref{eq:finalKCM}) are known to have activated dynamics \cite{Ritort}.  
This suggests that there may be alternative coarse graining
schemes that capture a connection between MD simulations and
KCM.  Nevertheless, to the best of our knowledge, we have demonstrated the first effective 
mapping of MD simulations onto a KCM.  The ``spins'' 
of our model correspond to regions of high or low MSD per particle and hence are in 
the spirit of the ``mobility field''\cite{PNAS}.   

Given that the mapping we exhibit does not allow us to obtain the fragile glass
behaviour of the original model from our KCM, it is interesting to ask what
physics the spins that we map to are sensitive to.  We believe they are sensitive to
slow structural relaxation that proceeds on time scales even longer than $t_\alpha$.  This appears
to be consistent with recent work by Szamel and Flenner \cite{Szamel}
 that suggests that continuing relaxations beyond $t_\alpha$ lead to 
the timescale for the onset of Fickian 
diffusion to be an order of magnitude longer than $t_\alpha$ and grow faster 
than $t_\alpha$ at low temperatures. The same authors also proposed a non-gaussian
parameter which differs from that conventionally used in studies of glass formers \cite{Flenner}.
This non-gaussian parameter has a peak at later times than the conventional
one, at timescales of order $t_\alpha$, where it appears that heterogeneity in 
the distribution of particle displacements is maximal.  This might be related
to why a ``spin'' definition based on mean square displacement, as we use here,
is most sensitive to physics on the timescale of $t_\alpha$.  This suggests
that in order to make a mapping to a KCM more in line with naive expectations,
one should use quantities that have maximum contrast on timescales which are considerably
shorter than $t_\alpha$.  Such an approach may be a viable way
to construct effective theories of glassy dynamics in this and in other systems.
We have demonstrated a particular numerical coarse-graining procedure, and we hope that 
our results may help point the way to analytic approaches to connect microscopic
models of glasses to KCMs, and further insight  into the glass problem.

\section{Acknowledgements}

Calculations were performed on Westgrid.  The authors thank 
Horacio Castillo, Claudio Chamon, and Mike 
Plischke for discussions, and David Reichman for critical
comments on the manuscript.  The authors also thank the anonymous referees for their
comments. This work was supported by NSERC.


\begin{thebibliography}{99}

\bibitem{Classic}  T. R. Kirkpatrick and P. G. Wolynes, Phys. Rev. A {\bf 35}, 3072 (1987);
         Phys. Rev. B {\bf 36}, 8552 (1987);
         T. R. Kirkpatrick and D. Thirumalai, Phys. Rev. Lett. {\bf 58}, 2091 (1987);
         C. A. Angell, J. Phys. Cond. Mat. {\bf 12}, 6463 (2000);
         G. Tarjus, {\it et al.}, J. Phys. Cond. Mat. {\bf 12}, 6497 (2000); 
         P. G. Debenedetti and F. H. Stillinger, Nature {\bf 410}, 259 (2001).

\bibitem{Sillescu} H. Sillescu, J. Non-Cryst. Solids {\bf 243}, 81 (1999).

\bibitem{Ediger} M. D. Ediger, Annu. Rev. Phys. Chem. {\bf 51}, 99 (2000).

\bibitem{Glotzer} S. C. Glotzer, J. Non-Cryst. Solids {\bf 274}, 342 (2000).

\bibitem{Richert} R. Richert, J. Phys. Cond. Mat. {\bf 14}, R703 (2002).

\bibitem{Castillo} H. E. Castillo, C. Chamon, L. F. Cugliandolo, 
            and M. P. Kennett, Phys. Rev. Lett. {\bf 88}, 237201 (2002); 
            C. Chamon, M. P. Kennett, H. E. Castillo, and L. F. Cugliandolo, 
            {\it ibid.} {\bf 89}, 217201 (2002);
            H. E. Castillo, C. Chamon, L. F. Cugliandolo, J. L. Iguain, and 
            M. P. Kennett, Phys. Rev. B {\bf 68}, 134442 (2003);
            C. Chamon, P. Charbonneau, L. F. Cugliandolo, D. R. Reichman, and  M. Sellitto, 
            J. Chem. Phys. {\bf 121}, 10120 (2004);
            H. E. Castillo and A. Parsaeian, Nature Phys. {\bf 3}, 26 (2007).

\bibitem{DHexp} E. Vidal-Russell and N. E. Israeloff, Nature {\bf 408}, 695 (2000);
        L. A. Deschenes and D. A. Vanden Bout, Science {\bf 292}, 255 (2001);
        S. A. Reinsberg, {\it et al.}, J. Chem. Phys. {\bf 114}, 7299 (2001);
        E. R. Weeks and D. A. Weitz, Phys. Rev. Lett. {\bf 89}, 095704 (2002);
        P. S. Crider and N. E. Israeloff, Nano. Lett. {\bf 6}, 887 (2006).

\bibitem{Kob} W. Kob, C. Donati, S. J. Plimpton, P. H. Poole, 
           and S. C. Glotzer, Phys. Rev. Lett. {\bf 79}, 2827 (1997).

\bibitem{Donati1} C. Donati, J. F. Douglas, W. Kob, S. J. Plimpton, P. H. Poole, 
                 and S. C. Glotzer, Phys. Rev. Lett. {\bf 80}, 2338 (1998).

\bibitem{Donati2} C. Donati, S. C. Glotzer, P. H. Poole, W. Kob, 
              and S. J. Plimpton, Phys. Rev. E {\bf 60}, 3107 (1999).

\bibitem{Adam} G. Adam and J. H. Gibbs, J. Chem. Phys. {\bf 43}, 139 (1965).

\bibitem{HarrowellDH} S. Butler and P. Harrowell, J. Chem. Phys. {\bf 95}, 4454 (1991);
          {\it ibid}, {\bf 95}, 4466 (1991); P. Harrowell, Phys. Rev. E {\bf 48}, 4359 (1993); 
          D. N. Perera and P. Harrowell, {\it ibid.} {\bf 54}, 1652 (1996).

\bibitem{Yamamoto} R. Yamamoto and A. Onuki, Phys. Rev. Lett. {\bf 81}, 4915 (1998);
       Phys. Rev. E {\bf 58}, 3515 (1998).

\bibitem{Andersen} H. C. Andersen, Proc. Nat. Acad. Sci. {\bf 102}, 6686 (2005).

\bibitem{Perera2} D. N. Perera and P. Harrowell, J. Non-Cryst.
                 Solids {\bf 235}, 314 (1998).

\bibitem{Gebremichael1} Y. Gebremichael, {\it et al.}, J. Chem. Phys.
              {\bf 120}, 4415 (2004).

\bibitem{Stevenson}  J. D. Stevenson, {\it et al.},  Nature Phys. {\bf 2}, 268 (2006).

\bibitem{Kegel-Weeks} W. K. Kegel and A. van Blaarderen, Science {\bf 287}, 290 (2000);
                E. R. Weeks, {\it et al.}, Science {\bf 287}, 627 (2000).

\bibitem{Ritort} F. Ritort and P. Sollich, Adv. Phys. {\bf 52}, 219 (2003).

\bibitem{FA} G. H. Fredrickson and H. C. Andersen, Phys. Rev. Lett. {\bf 53}, 1244 (1984);
             J. Chem. Phys. {\bf 83}, 5822 (1985).

\bibitem{Graham} I. S. Graham, L. Pich\'{e} and M. Grant, Phys. Rev. E {\bf 55}, 2132 (1997).

\bibitem{CG} J. P. Garrahan and D. Chandler, Phys. Rev. Lett. {\bf 89}, 035704 (2002);
             L. Berthier and J. P. Garrahan, J. Chem. Phys. {\bf 119}, 4367 (2003);
             L. Berthier, Phys. Rev. Lett. {\bf 91}, 055701 (2003); 
             L. Berthier and J. P. Garrahan, Phys. Rev. E {\bf 68}, 041201 (2003).

\bibitem{PNAS} J. P. Garrahan and D. Chandler, 
                Proc. Nat. Acad. Sci. {\bf 100}, 9710 (2003).

\bibitem{LMSBG} S. L\'{e}onard, {\it et al.}, cond-mat/0703164.

\bibitem{Evans} P. Sollich and M. R. Evans, Phys. Rev. Lett. {\bf 83}, 3238 (1999).

\bibitem{ARao} C. A. Angell and K. J. Rao, J. Chem. Phys. {\bf 57}, 470 (1972).

\bibitem{Vogel} M. Vogel and S. C. Glotzer, Phys. Rev. Lett. {\bf 92}, 255901 (2004);
                M. N. J. Bergroth, {\it et al.}, J. Phys. Chem. B
               {\bf 109}, 6748 (2005).

\bibitem{Perera} D. N. Perera and P. Harrowell, Phys. Rev. Lett. {\bf 80}, 4446 (1998);
         {\it ibid.} {\bf 81}, 120 (1998); Phys. Rev. E {\bf 59}, 5721 (1999); 
         J. Chem. Phys {\bf 111}, 5441 (1999).

\bibitem{AWC1} A. Widmer-Cooper, P. Harrowell, and H. Fynewever, 
              Phys. Rev. Lett. {\bf 93}, 135701 (2004).

\bibitem{footnote} We choose a square lattice for simplicity.

\bibitem{Dasgupta} C. Dasgupta, A. V. Indrani, S. Ramaswami, and M. K. Phani, Europhys. Lett.
   {\bf 15}, 307 (1991).

\bibitem{Donati3} C. Benneman, C. Donati, J. Baschnagle, and S. C. Glotzer, Nature {\bf 399}, 246
 (1999); C. Donati, S. C. Glotzer, and P. H. Poole, Phys. Rev. Lett. {\bf 82}, 5064 (1999);
 C. Donati, S. Franz, G. Parisi, and S. C. Glotzer, J. Non-Cryst. Solids {\bf 307}, 215 (2002). 

\bibitem{Doliwa} B. Doliwa and A. Heuer, Phys. Rev. E {\bf 61}, 6898 (2000).

\bibitem{FranzParisi} S. Franz and G. Parisi, J. Phys. Cond. Mat. {\bf 12}, 6335 (2000).

\bibitem{Glotzer2} S. C. Glotzer, V. V. Novikov, and T. B. Schroder, J. Chem. Phys. {\bf 112},
     509 (2000).

\bibitem{BB} G. Biroli and J.-P. Bouchaud, Europhys. Lett. {\bf 67}, 21 (2004).

\bibitem{Toninelli} C. Toninelli, M. Wyart, L. Berthier, G. Biroli, and J.-P. Bouchaud,
      Phys. Rev. E {\bf 71}, 041505 (2005).

\bibitem{Berthier} L. Berthier, G. Biroli, J.-P. Bouchaud, L. Cipelletti, D. El Masri,
 D. L'H\^{o}te, F. Ladieu, and M. Pierno, Science {\bf 310}, 1797 (2005); L. Berthier,
G. Biroli, J.-P. Bouchaud, W. Kob K. Miyazaki, and D. Reichman, J. Chem. Phys. {\bf 126}, 184503
(2007); {\it ibid} 184504 (2007).

\bibitem{Szamel} G. Szamel and E. Flenner, Phys. Rev. E {\bf 73}, 011504 (2006).

\bibitem{Flenner} E. Flenner and G. Szamel, Phys. Rev. E {\bf 72}, 011205 (2005).

\end{thebibliography}
\end{document}